\documentclass[english,aps,manuscript,preprintnumbers,superscriptaddress,eqsecnum,nofootinbib]{revtex4}
\usepackage[T1]{fontenc}
\usepackage[utf8]{luainputenc}
\usepackage{array}
\usepackage{appendix}
\usepackage{multirow}
\usepackage{amsmath}
\allowdisplaybreaks[4]
\usepackage{amssymb}
\usepackage{float}

\usepackage{wasysym}
\usepackage{graphicx}
\usepackage{esint}
\usepackage[marginal]{footmisc}

\numberwithin{equation}{section}
\DeclareMathSizes{12}{12}{7}{7}
\renewcommand{\theequation}{\arabic{section}.\arabic{equation}}

\renewcommand\thesection{\arabic{section}}

\renewcommand{\baselinestretch}{1.1}
\makeatletter

\@ifundefined{textcolor}{}
{%
 \definecolor{BLACK}{gray}{0}
 \definecolor{WHITE}{gray}{1}
 \definecolor{RED}{rgb}{1,0,0}
 \definecolor{GREEN}{rgb}{0,1,0}
 \definecolor{BLUE}{rgb}{0,0,1}
 \definecolor{CYAN}{cmyk}{1,0,0,0}
 \definecolor{MAGENTA}{cmyk}{0,1,0,0}
 \definecolor{YELLOW}{cmyk}{0,0,1,0}
}
\numberwithin{figure}{section}

\makeatother

\usepackage{babel}
\begin{document}

\title{The Leptophilic Dark Matter in the Sun: the Minimum Testable Mass }

\author{Zheng-Liang Liang}

\email{liangzl@itp.ac.cn}
\affiliation{Institute of Applied Physics and Computational Mathematics\\Beijing, 100088, China}

\author{Yi-Lei Tang}
\email{tangyilei@kias.re.kr}
\affiliation{School of Physics, KIAS,\\85 Hoegiro, Seoul 02455, Republic of Korea}

\author{Zi-Qing Yang}
\email{ziqingyang@google.com}
\affiliation{Big Data Research, IFLYTEK\\Wangjiang Road West 666\#, Hefei, 230088, China}

\vspace{5cm}

\begin{abstract}
The physics of the solar dark matter (DM) that are captured and thermalise
through the DM-nucleon interaction has been extensively studied.
In this work, we consider the leptophilic DM scenario where the DM
particles interact exclusively with the electrons through the axial-vector
coupling. We investigate relevant phenomenologies in the Sun, including
its capture, evaporation and thermalisation, and we calculate the
equilibrium distribution using the Monte Carlo methods, rather than
adopting a semi-analytic approximation. Based on the analysis, we
then determine the minimum testable mass for which the DM-electron
coupling strength can be probed via the neutrino observation. Compared to the case of the DM-nucleon interaction, it turns out
that minimum detectable mass of the DM-electron interaction is roughly
1 GeV smaller, and a cross section about two orders of magnitude larger  is required for the saturation of  the annihilation signal. \vspace{3cm}

\end{abstract}
\maketitle

\section{\label{sec:Introduction}Introduction}

Several neutrino telescopes have been looking for the trace of the
Weakly Interacting Massive Particles (WIMPs)~\cite{2011ApJ...742...78T,Aartsen:2012kia,Adrian-Martinez:2013ayv,Avrorin:2014swy},
a generic kind of candidate for the Dark Matter (DM), from the Sun.
This is based on the picture that the Galactic WIMPs collides with
nuclei in the Sun as they pass by the solar neighbourhood, gradually
sinking into the solar core after subsequent collisions, and end up
annihilating into primary or secondary neutrinos that escape the environment
of the dense plasma in the Sun, so to be observed by the terrestrial
neutrino detectors.

The neutrino flux at the detector location is related to the solar
DM annihilation through the following schematic relation:
\begin{equation}
\frac{d\Phi_{\nu}}{dE_{\nu}}=\frac{\Gamma_{A}}{4\pi d_{\odot}^{2}}\frac{dN_{\nu}}{dE_{\nu}},
\end{equation}
where $d_{\odot}$ is the Sun-Earth distance, $d\Phi_{\nu}/dE_{\nu}$
and $dN_{\nu}/dE_{\nu}$ represent the neutrino differential flux
at the Earth and the neutrino energy spectrum per DM annihilation
event in the Sun, respectively. The total annihilation rate $\Gamma_{A}$
can be expressed in terms of the number of the trapped DM particles
$N_{\chi}$:
\begin{eqnarray}
\varGamma_{A} & = & \frac{1}{2}A_{\odot}N_{\chi}^{2},
\end{eqnarray}
where $A_{\odot}$ denotes twice the annihilation rate of a pair of
DM particles. The evolution of the solar DM number $N_{\chi}$ is
depicted with the following equation:
\begin{equation}
\frac{dN_{\chi}}{dt}=C_{\odot}-E_{\odot}N_{\chi}-A_{\odot}N_{\chi}^{2},\label{eq:master equation-1}
\end{equation}
which involves the DM capture (evaporation) rate $C_{\odot}\,\left(E_{\odot}\right)$
by scattering off atomic nuclei in the Sun, as well as the annihilation
rate $A_{\odot}$. Eq.~(\ref{eq:master equation-1}) has an analytic
solution
\begin{equation}
N_{\chi}\left(t\right)=\frac{C_{\odot}\,\tanh\left(t/\tau_{\mathrm{e}}\right)}{\tau_{\mathrm{e}}^{-1}+\left(E_{\odot}/2\right)\tanh\left(t/\tau_{\mathrm{e}}\right)},\label{eq:DMnumber}
\end{equation}
with

\begin{equation}
\tau_{\mathrm{e}}=\left(C_{\odot}\, A_{\odot}+E_{\odot}^{2}/4\right)^{-1/2}
\end{equation}
 the time scale for the capture, evaporation and annihilation processes
to equilibrate. Once the equilibrium is reached at the present day,
$i.e.$, $\tanh\left(t_{\odot}/\tau_{\mathrm{e}}\right)\simeq1$,
with $t_{\odot}=4.5\times10^{9}\,\mathrm{yr}$ being the solar age,
the annihilation output $\varGamma_{A}$ also reaches its maximum
value. Depending on the ratio $E_{\odot}^{2}/\left(C_{\odot}A_{\odot}\right)$,
or the DM mass $m_{\chi}$, such equilibrium can be categorized into
two different scenarios: (1) $E_{\odot}^{2}/\left(C_{\odot}A_{\odot}\right)\ll1$,
that's when the evaporation effect can be neglected and the equilibrium
is between annihilation and  capture. In this case, $N_{\chi}\simeq C_{\odot}/E_{\odot}
 $ and hence $\Gamma_{A}\simeq C_{\odot}/2$,
so we can either determine or constrain the strength of the DM-nucleon
interaction from solar neutrino observation; (2) $E_{\odot}^{2}/\left(C_{\odot}A_{\odot}\right)\gg1$,
under this circumstance evaporation overwhelms annihilation for the
DM depletion, and the balance between evaporation and capture
yields $N_{\chi}\simeq\sqrt{C_{\odot}/A_{\odot}}$ and  $\Gamma_{A}\simeq A_{\odot}C_{\odot}^{2}/\left(2\, E_{\odot}^{2}\right)$,
which not only implies a heavy suppression of the neutrino flux, but
also prevents us from drawing the coupling strength of the DM-nucleon
interaction from the possible observed signals.

While relevant phenomenology associated with the DM-nucleon interaction
have been studied extensively in literature, the tempting possibility
that the DM particles couple exclusively to leptons, the so-called
leptophilic scenario, has aroused wide interest in the community~\cite{Bernabei:2007gr,Fox:2008kb,Kopp:2009et,Feldstein:2010su,Essig:2011nj,Dev:2013hka,Foot:2014xwa,Lee:2015qva,Roberts:2016xfw}.
However, even for a broad range of leptophilic DM models, it turns
out that the the effective DM-nucleon cross section arising from the
loop-induced DM-quark interaction competes with or overwhelms that of the
DM-electron interaction~\cite{Kopp:2009et}. A notable exception is that
the DM particle interacts with electron through the axial-vector coupling,
a case in which the loop-induced contribution vanishes.

In this work, we will investigate some interesting phenomena of the
leptophilic DM trapped in the Sun. Specifically, we will explore the
minimum testable mass through neutrino observation for the scenario
where the DM particle couples exclusively to electron. The minimum
detectable solar DM mass is determined by the parametric relations
between capture, evaporation and annihilation, as has been extensively
studied in the context of the DM-nucleon coupling scenario~\cite{Spergel:1984re,Griest:1986yu,Nauenberg:1986em,Gould:1987ju,Gould:1989hm,Gould:1991hx,Nussinov:2009ft,Busoni:2013kaa,Liang2014,Catena:2015uha,Vincent:2015gqa,Blennow:2015hzp,Dev:2015isx,Kouvaris:2015nsa,Vincent:2016dcp,Liang:2016yjf,Baum:2016oow,Garani:2017jcj,Smolinsky:2017fvb,Busoni:2017mhe,Widmark:2017yvd}.
But considering that the medium of the ionised electrons is much softer
than that of the nuclei (suppressed by a factor $\sqrt{m_{e}/m_{N}}$ in terms of thermally averaged momentum,
with the electron (nucleus) mass $m_{e}$ ($m_{N}$)), the minimum
detectable leptophilic DM mass is expected to be smaller accordingly,
due to the less energetic collisions that would prevent the buildup
of the solar DM through evaporation. While  quantitative analyses
on this issue have been discussed in ref.~\cite{Garani:2017jcj},
where equilibrium distribution of leptophilic DM was described phenomenologically
with a semi-analytic approximation, in this paper, we will pursue
an accurate evaluation of the distribution with a Monte Carlo method
adopted in refs.~\cite{Gould:1987ju,Liang:2016yjf,Blennow:2018xwu}, in an effort
to provide a more precise description of the leptophilic DM in the
Sun. Interestingly, we find that the simulated distribution is remarkably suppressed at the high velocity end when compared to the truncated Boltzmann approximation adopted in ref.~\cite{Garani:2017jcj}, and results in an evaporation rate roughly 4 orders of magnitude smaller. As a consequence, such difference translates to an evaporation mass around 1~GeV smaller.

This paper is organised as follows. In sec.~\ref{sec:Distribution-and-evolution},
we will take a brief review on the theoretical ground for the capture,
evaporation, and annihilation of the leptophilic DM in the Sun, and
put these formulas into numerical computation. Main results, along
with relevant analyses and discussions, are provided in sec.~\ref{sec:MinimumTestableMass}.

\section{\label{sec:Distribution-and-evolution}Distribution and evolution
of solar DM}

In this section we will discuss the distribution and evolution of
the solar DM. An accurate description of the distribution of the captured
DM is crucial for the evaluation of evaporation and annihilation rate,
and together with capture rate, they determine the evolution of the
solar DM population. We obtain the solar DM distribution by solving
the Boltzmann equation in a numerical manner. Now we delve into the
details.

\subsection{\label{sub:capture}capture of the dark matter by solar electrons}

The buildup of the solar DM population begins with the capture of
the Galactic DM particles. There is the possibility that the free-streaming
DM particles will be gravitationally pulled inside the Sun and scattered
by electrons therein to velocities lower than the local escape velocity,
so to be captured. The standard procedure for evaluating the DM capture
rate $\widetilde{C_{\odot}}$ has been well established in the literature~\cite{Gould:1987ir,Gould:1987ww,Gould:1991hx}.
After a small modification to replace nuclei with electrons, the capture
rate of the DM particle by solar electrons can be expressed as
\begin{eqnarray}
\widetilde{C_{\odot}} & = & \frac{\rho_{\chi}}{m_{\chi}}\int_{0}^{R_{\odot}}4\pi r^{2}\mathrm{d}r\,\int\frac{w}{u}\, g_{\chi}\left(\mathbf{u}\right)\mathrm{d}^{3}u\int_{0}^{v_{\mathrm{esc}}}R_{e}\left(w\rightarrow v\right)\mathrm{d}v,\label{eq:CapRate}
\end{eqnarray}
where $R_{\odot}$ is the radius of the Sun, $m_{\chi}$ is the
mass of the DM particle, $\rho_{\chi}$
and $g_{\chi}\left(\mathbf{u}\right)$ are the DM density in the solar
neighborhood and the velocity distribution in the solar rest frame,
respectively. In calculation, we use $\rho_{\chi}=0.3\,\mathrm{GeV\cdot cm^{-3}}$ and model the velocity distribution as
a Maxwellian form in the rest frame of the Galactic centre, with the dispersion velocity $v_{0}=220\,\mathrm{km\cdot s}^{-1}$
and a truncation at the Galactic escape velocity of $544\,\mathrm{km\cdot s}^{-1}$.
$w=\sqrt{v_{\mathrm{esc}}^{2}+u^{2}}$ connects the velocity outside
the solar influence sphere, $u$, and the one accelerated by the gravitational
pull at the radius $r$, with the local escape velocity $v_{\mathrm{esc}}$.
The quantity $R_{e}^{-}\left(w\rightarrow v\right)$ represents the
differential event rate of a DM particle with initial velocity $w$
down-scattered to final smaller one $v$ by solar electrons in unit
volume as the following,
\begin{eqnarray}
R_{e}^{-}\left(w\rightarrow v\right) & = & n_{e}\left\langle \frac{\mathrm{d}\sigma_{\chi e}}{\mathrm{d}v}\left|\mathbf{w}-\mathbf{u}_{e}\right|\right\rangle \nonumber \\
 & = & n_{e}\int f_{e}\left(\mathbf{u}_{e}\right)\frac{\mathrm{d}\sigma_{\chi e}}{\mathrm{d}v}\,\left|\mathbf{w}-\mathbf{u}_{e}\right|\mathrm{d}^{3}u_{e},\label{eq:Re}
\end{eqnarray}
where the differential cross section for the DM-electron system $\mathrm{d}\sigma_{\chi e}/\mathrm{d}v$
depends on their relative velocity $\mathbf{w}-\mathbf{u}_{e}$, and
$\left\langle \cdots\right\rangle $ denotes the average over the
thermal velocity distribution of solar electrons. $n_{e}$ is the
local electron number density. The Maxwellian distribution $f_{e}\left(\mathbf{u}_{e}\right)$
is written as
\begin{eqnarray}
f_{e}\left(\mathbf{u}_{e}\right) & = & \left(\sqrt{\pi}u_{0}\right)^{-3}\exp\left(-\frac{u_{e}^{2}}{u_{0}^{2}}\right),
\end{eqnarray}
with $u_{0}=$ $\sqrt{2\ T_{\odot}/m_{e}}$.
{$T_{\odot}\left(r\right)$ is the local temperature
at radius $r$. }Hence the event rate eq.~(\ref{eq:Re}) can be further
expressed explicitly in the following analytic form,
\begin{eqnarray}
R_{e}^{-}\left(w\rightarrow v\right) & = & \frac{n_{e}\,\sigma_{\chi e}}{4\,\eta}\left(\eta^{+}\right)^{2}\frac{v}{w}\left\{ \mathrm{erf}\left[\frac{\left(\eta^{-}w-\eta^{+}v\right)}{2u_{0}},\,\frac{\left(\eta^{-}w+\eta^{+}v\right)}{2u_{0}}\right]\right.\nonumber \\
\nonumber \\
 &  & \left.+\exp\left[\eta\left(w^{2}-v^{2}\right)/u_{0}^{2}\right]\,\mathrm{erf}\left[\frac{\left(\eta^{+}w-\eta^{-}v\right)}{2u_{0}},\,\frac{\left(\eta^{+}w+\eta^{-}v\right)}{2u_{0}}\right]\right\} ,
\end{eqnarray}
where $\eta^{\pm}\equiv\eta\pm1=m_{\chi}/m_{e}+1$, and $\mathrm{erf}\left(a,\, b\right)\equiv\mathrm{erf}\left(b\right)-\mathrm{erf}\left(a\right)$.

\subsection{\label{sub:sec.3.a}relaxation and distribution of the solar DM}

In order to determine the distribution of the trapped DM, in this
work we adopt the numerical method outlined in ref.~\cite{Gould:1987ju},
which is in essence equivalent to solving the Boltzmann equation.
The benefits are two-fold: first, this method is able to describe
the high end of the velocity distribution of the solar DM, which is
a prerequisite of an accurate evaluation of evaporation rate; second,
this method can also provide a detailed description of the relaxation
process of the newly captured DM particles. The second advantage is
especially important because in contrast to the nucleus capture scenario,
the marginally captured DM particles by solar electrons are prone
to be ejected back to deep space before their distribution reach the
final equilibrium. The leakage due to the evaporation over the relaxation
process needs to be carefully quantified.

Here we take a brief introduction to the methodology. Our discussion
is based on the assumption that the accumulated DM population does
$not$ bring any significant impact on the solar structure, $i.e.$,
the Sun as a heat reservoir is also modeled as the background for the residing
DM particles. We keep track of a small portion of DM particles since
their capture until they finally integrate into the rest already in
equilibrium. The Boltzmann equation is linear due to the absence of
the DM self-interaction\footnote[1]{\renewcommand{\baselinestretch}{1}\selectfont Since only a small increment of solar DM particles is under investigation, annihilation effects can be all attributed to solar DM particles already in equilibrium. A detailed discussion is arranged at the end of sec.~\ref{sec:MinimumTestableMass}}, and can be further simplified as the following
\textit{master equation }if expressed with a convenient choice of
parameters $E$ (total energy per  mass) and $L$ (angular momentum
per mass) \cite{Gould:1987ju}:
\begin{eqnarray}
\frac{\mathrm{d}f\left(E,\, L,\, t\right)}{\mathrm{d}t} & = & -f\left(E,\, L,\, t\right)\sum_{E',L'}S\left(E,\, L;\, E',\, L'\right)+\sum_{E',L'}f\left(E',\, L',\, t\right)S\left(E',\, L';\, E,\, L\right)\nonumber \\
f\left(E,\, L,\,0\right) & = & f_{\mathrm{cap}}\left(E,\, L\right),\label{eq:master equation}
\end{eqnarray}
where $f\left(E,\, L,\, t\right)$ is distribution function at time
$t$, and $S\left(E,\, L;\, E',\, L'\right)$ represents the scattering
matrix element for transition process $\left(E,\, L\right)\rightarrow\left(E',\, L'\right)$.
In practice the parameters $E$ and $L$ are nondimensionalised in
terms of an energy reference value $GM_{\odot}/R_{\odot}$, and an
angular momentum value $\left(GM_{\odot}R_{\odot}\right)^{1/2}$.
These values are constructed from a length unit, namely the solar
radius $R_{\odot}=6.955\times10^{5}\,\mathrm{km}$, and a time unit
$\left(GM_{\odot}/R_{\odot}^{3}\right)^{-1/2}=1.596\times10^{3}\,\mathrm{s}$,
with the Newton's constant $G$ and the solar mass $M_{\odot}=1.988\times10^{30}\,\mathrm{kg}$.

\begin{figure}
\centering{}\includegraphics[scale=0.72]{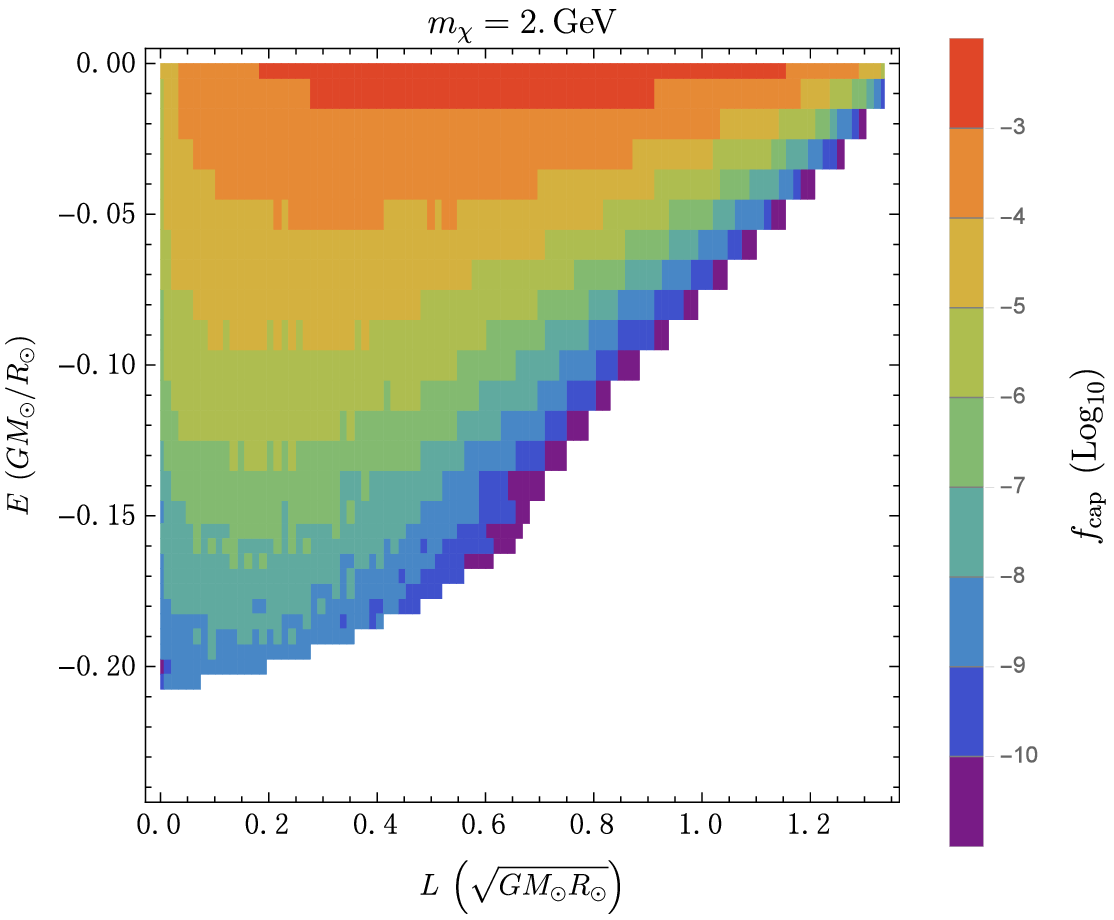}$\quad$\includegraphics[scale=0.68]{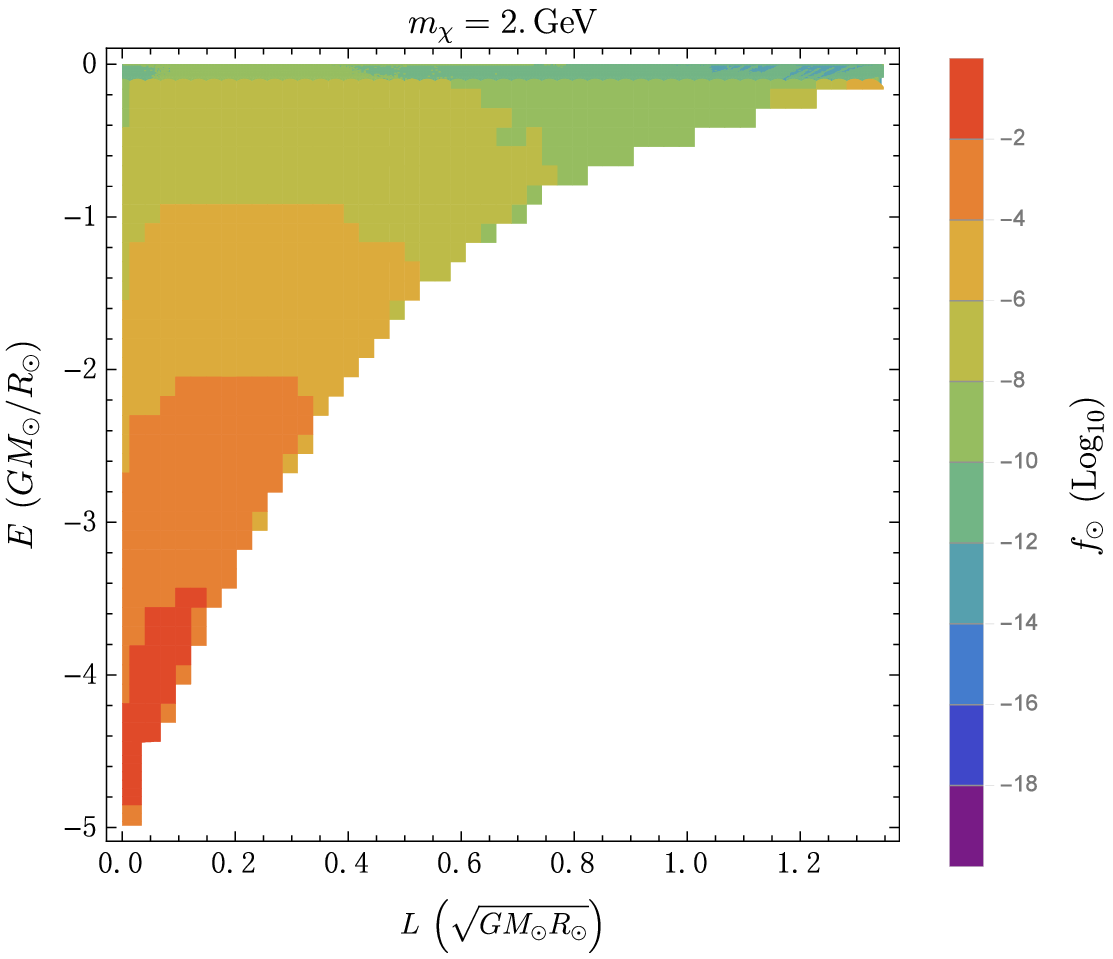}\protect\caption{\label{fig:EL_Dist} Initial distribution $f_{\mathrm{cap}}\left(E,\, L\right)$
(\textit{left}) and the limit distribution $f_{\odot}$ (\textit{right})
of a $2$ GeV DM particle, respectively. The energy $E$ and angular
momentum $L$ are nondimensionalised in units of $GM_{\odot}/R_{\odot}$
and $\left(GM_{\odot}R_{\odot}\right)^{1/2}$, respectively. Only
the coloured parameter region is allowed for bound orbits in the $right$
panel. See text for details.}
\end{figure}

As mentioned in ref.~\cite{Liang:2016yjf}, in the optical thin limit and for a large enough time step $\Delta t$,
the probability for a collision in simulation is insensitive to the
starting position in the periodic orbit defined by $E$ and $L$,
so these two parameters are sufficient for the representation of DM
states bound to the Sun. The initial distribution $f_{\mathrm{cap}}\left(E,\, L\right)$
is obtained by assuming the DM particles captured after first collision
with electrons do not deviate much from their incident directions.
Consequently, if $\vartheta$ is defined as the angle between the
trajectory of scattered DM particle and radial direction from the
centre of the Sun, the probability distribution scale proportionally
with differential $\mathrm{d}\left(\cos\vartheta\right)$ in each
spherical layer $4\pi r^{2}\mathrm{d}r$, with $0\leq\vartheta\leq\frac{\pi}{2}$.
For illustration, we present the initial distribution $f_{\mathrm{cap}}\left(E,\, L\right)$
for a $2$ GeV DM particle in the left panel of fig.~\ref{fig:EL_Dist}.

It should be noted that compared with nucleus, the electron thermal
momentum is suppressed by $\sqrt{m_{e}/m_{N}}$, which suggests that the DM particle can only marginally
fall into, and easily escape from the solar gravitational well. Therefore
the evaporation effect is taken into account in eq.~(\ref{eq:master equation}),
at variance with the approach adopted in ref.~\cite{Liang:2016yjf},
where only gravitationally bound states are involved in the simulation
of relaxation process. This modification is necessary considering
that the leakage due to evaporation may no longer be neglected over
the relaxation process. Since all bound states are connected and the
evaporated DM particles are not anticipated to be trapped again, in
simulation we allocate one state to account for the escape state that
is corresponding to the \textit{absorbing} state in the context of
Markov process, whereas all bound states are corresponding to the
\textit{transient} states. For long enough time, evaporation will
deplete all the DM particles participating in simulation, but before
that a steady normalised distribution among the survival DM particles
is expected to be reached (for which we provide a proof in Appendix~\ref{sec:appendixb}).
We evolve eq.~(\ref{eq:master equation}) with the discrete time
step $\Delta t$ until $f\left(E,\, L\right)$ converges to this limiting
distribution $f_{\odot}\left(E,\, L\right)$, and other physical details
of the relaxation can be recorded at the same time. The equilibrium
distribution $f_{\odot}\left(E,\, L\right)$ is shown in the right panel of  fig.~\ref{fig:EL_Dist}.

\begin{figure}
\begin{centering}
\includegraphics[scale=0.6]{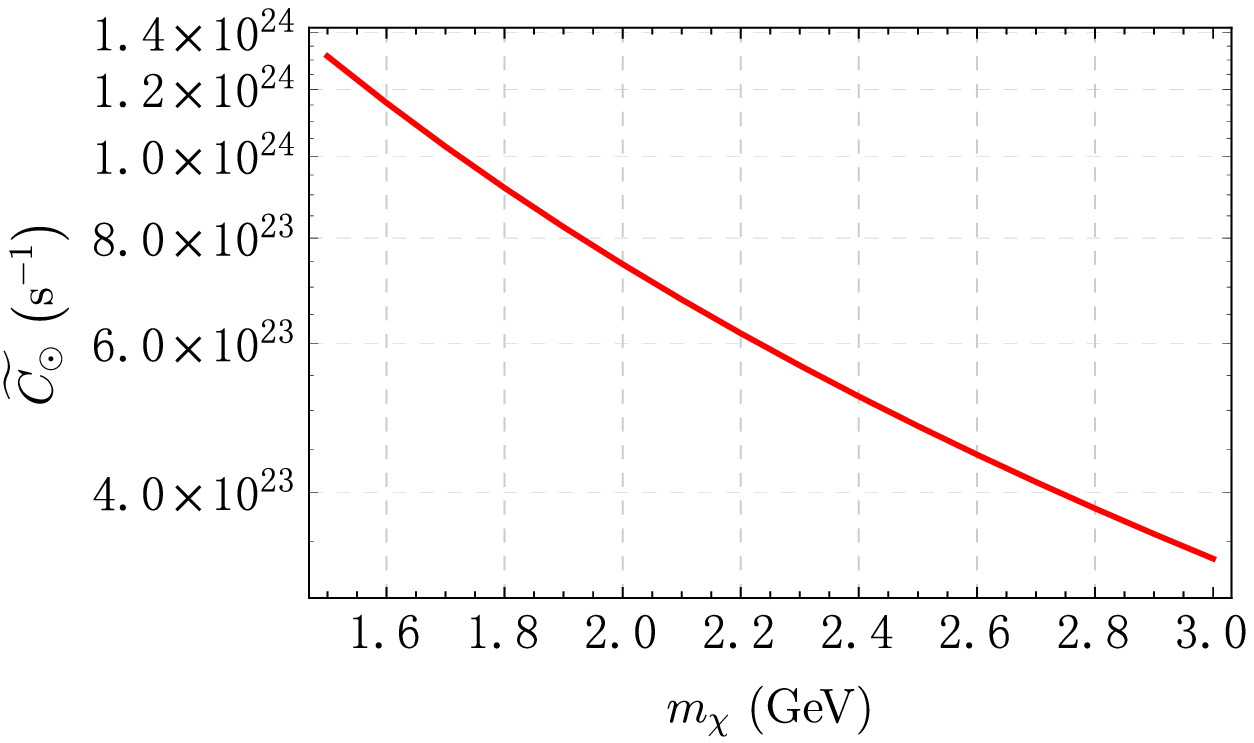}$\quad$\includegraphics[scale=0.56]{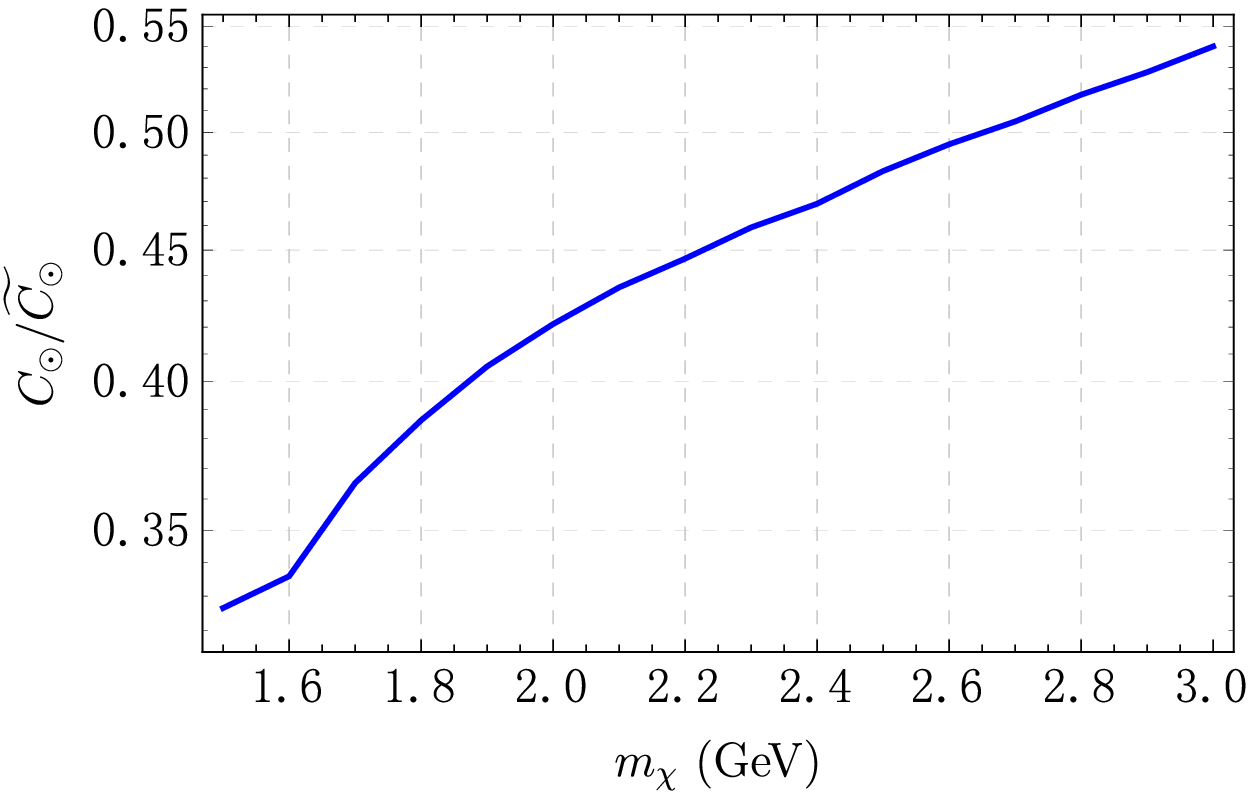}
\par\end{centering}

\protect\caption{\label{fig:captureRate}($left$) The original capture rate for the
DM mass range from $1.5$ to $3.0$ GeV, obtained from eq.~(\ref{eq:CapRate}).
($right$) Relevant ratio of the effective capture rate to the original
one, $C_{\odot}/\widetilde{C_{\odot}}$. See text for details. }
\end{figure}

In practice, we use
different levels of resolution to represent the bound states on the
$E$-$L$ plane. In order to accurately describe the transitions that
occur mostly near the escape state, the absolute value of energy
(angular momentum) parameter $E$ ($L$) is logarithmically (uniformly)
discretised in 40 (500) states from 0.01 (0) to 0.1 (1.35), and all bound states with energy above $-0.01$ are represented by $E=-0.01$, which is
equivalent to imposing a cutoff above 0.995 of the escape velocity on the distribution.
The second and the third regions are uniformly divided into $35\times50$
pieces on the $E$-$L$  plane from 0.1 (0) to 4.5 (1.35), and $10\times20$
pieces from 4.5 (0) to 5.04 (1.35), respectively. If the relaxation time
scale is verified to be much smaller than the Sun age $t_{\odot}$ (or $\tau_{\mathrm{e}}$),
the picture of instant thermalisation will keep unchanged except that
an effective capture rate $C_{\odot}$ should be introduced as the
original one $\widetilde{C_{\odot}}$, suppressed by the remaining proportion
over the relaxation process. For illustration, the capture rate $\widetilde{C_{\odot}}$
and the ratio between the two capture rates, $C_{\odot}/\widetilde{C_{\odot}}$,
are presented in fig.~\ref{fig:captureRate}, respectively.

The scattering matrix element $S\left(E,\, L;\, E',\, L'\right)$
is determined with Monte Carlo approach. Specifically, a large number
of DM random walk samples are generated and tallied in the fixed time
step $\Delta t$, which is required to be long enough to ensure that
the test particle receive substantial transfer momentum from solar
electrons. The periodic radial trajectory of the bound DM particle between
successive collisions  is numerically integrated
with the Standard Sun Model (SSM) GS98 \cite{Serenelli:2009yc} inside the Sun~($r\leq1$), and
is matched to analytic Keplerian orbit beyond the solar radius (if
any). Thus the $\left(i+1\right)$-th collision location and time
$t_{i+1}$ can be determined with the random renewal collision probability
$P_{\mathrm{c}}^{i}$ via
\begin{eqnarray}
P_{\mathrm{c}}^{i} & = & 1-\exp\left[-\int_{t_{i}}^{t_{i+1}}\lambda(\tau)\,\mathrm{d}\tau\right],\label{eq:collisionP}
\end{eqnarray}
where
\begin{eqnarray}
\lambda & = & n_{e}\left\langle \sigma_{\chi e}\left(\left|\mathbf{w}-\mathbf{u}_{e}\right|\right)\left|\mathbf{w}-\mathbf{u}_{e}\right|\right\rangle \nonumber \\
\nonumber \\
 & = & n_{e}\,\sigma_{\chi e}\left[\frac{u_{0}}{\sqrt{\pi}}\,\exp\left(-w^{2}/u_{0}^{2}\right)+\left(w+\frac{u_{0}^{2}}{2\, w}\right)\mathrm{erf}\left(\frac{w}{u_{0}}\right)\right]\label{eq:lambda}
\end{eqnarray}
is $implicitly$ dependent on the temporal parameter $\tau$ once
the DM trajectory is determined with the method mentioned above. By generating further random numbers that help pick out the colliding solar
electron's velocity, and the scattering angle in the centre-of-mass
(CM) frame, we then determine the outgoing state of the scattered
DM particle after a coordinate transformation back to the solar reference.

\subsection{\label{sub:evaporation}evaporation and annihilation}

\begin{figure}
\begin{centering}
\includegraphics[scale=0.7]{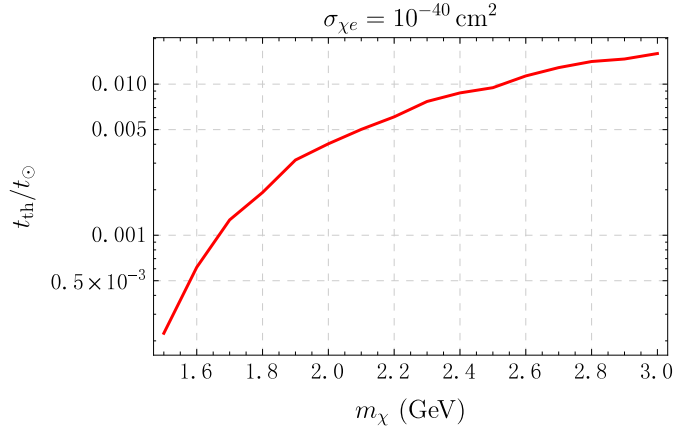}
\par\end{centering}
\protect\caption{\label{fig:thermTime}The thermalisation time of the captured solar DM particles in the simulation.}

\end{figure}

Given the distribution, both evaporation and annihilation rate of
the bound DM particle can be determined. The theoretical expression
of the evaporation rate differs with the capture rate only in the
way that the distribution of the incident DM particles is replaced
by the normalised distribution of the DM particles trapped in the
Sun, $f_{\odot}\left(r,w\right)$, and an up-scatter
event rate $R_{e}^{+}\left(w\rightarrow v\right)$ with $v>w$ is
introduced to account for the evaporation rather than $R_{e}^{-}\left(w\rightarrow v\right)$.
Thus the evaporation rate is expressed as
\begin{eqnarray}
E_{\odot} & = & \int_{0}^{R_{\odot}}\mathrm{d}r\,\int f_{\odot}\left(r,w\right)\,\mathrm{d}w\int_{v_{\mathrm{esc}}}^{+\infty}R_{e}^{+}\left(w\rightarrow v\right)\mathrm{d}v,
\end{eqnarray}
where
\begin{eqnarray}
R_{e}^{+}\left(w\rightarrow v\right) & = & \frac{n_{e}\,\sigma_{\chi e}}{4\,\eta}\left(\eta^{+}\right)^{2}\frac{v}{w}\left\{ \mathrm{erf}\left[\frac{\left(\eta^{+}v-\eta^{-}w\right)}{2u_{0}},\,\frac{\left(\eta^{+}v+\eta^{-}w\right)}{2u_{0}}\right]\right.\nonumber \\
 &  & +\left.\exp\left[\eta\left(w^{2}-v^{2}\right)/u_{0}^{2}\right]\,\mathrm{erf}\left[\frac{\left(\eta^{-}v-\eta^{+}w\right)}{2u_{0}},\,\frac{\left(\eta^{-}v+\eta^{+}w\right)}{2u_{0}}\right]\right\} .
\end{eqnarray}
Besides, the evaporation rate can also be determined from simulation straightforwardly.
Specifically speaking, the evaporation rate can also be constructed
by collecting all the inflow probability into the escape state in
each time step, at any instant time during the evolution of solar
DM.
\begin{figure}
\begin{centering}
\includegraphics[scale=0.85]{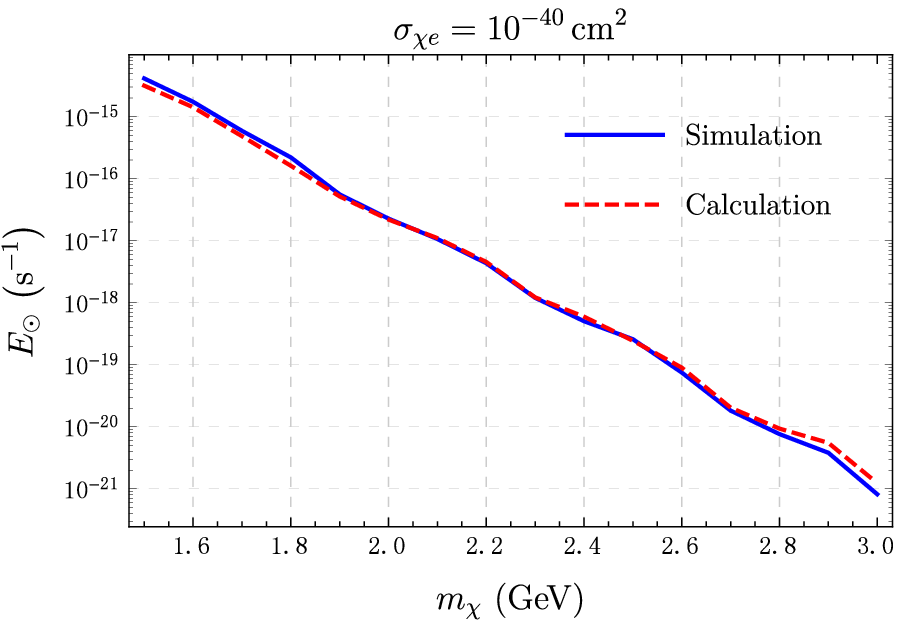}$\quad$\includegraphics[scale=0.63]{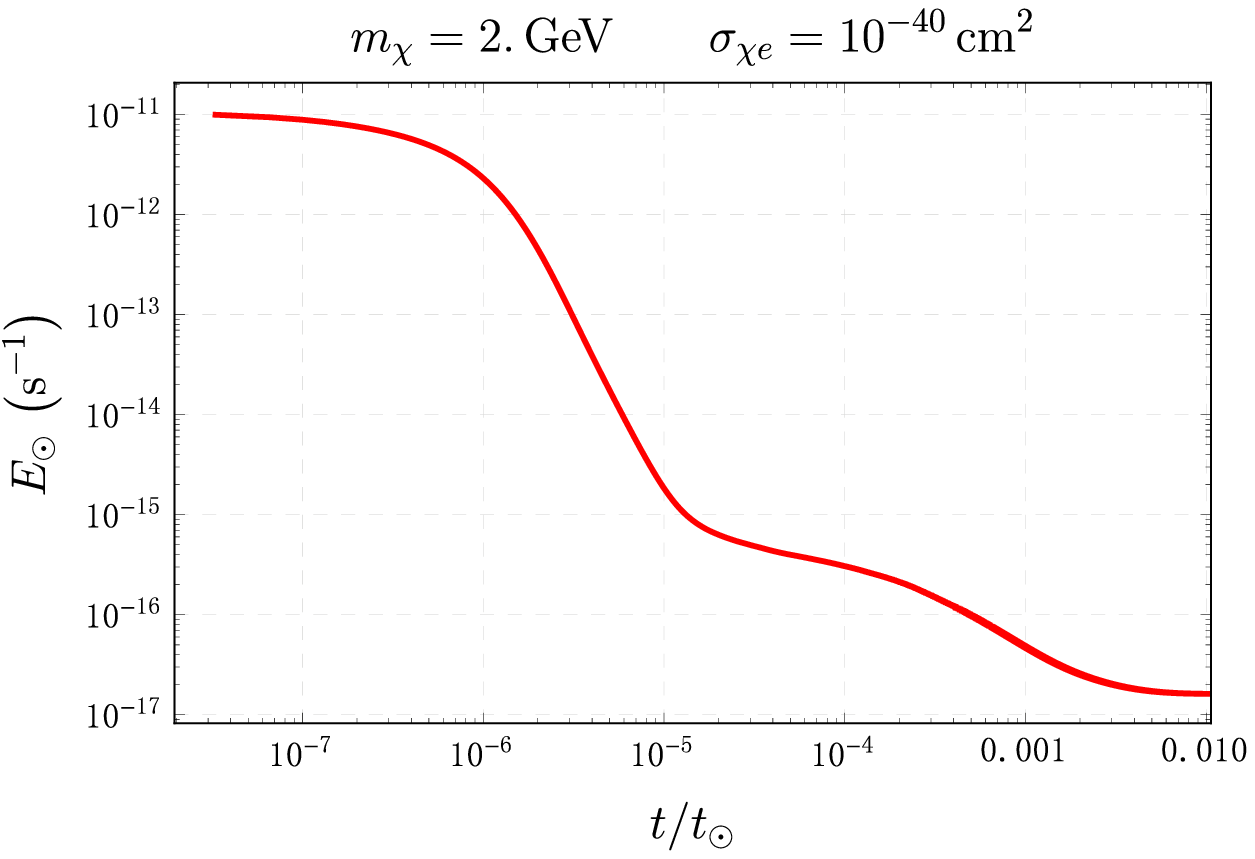}
\par\end{centering}

\protect\caption{\label{figEvaporation-rate} $Left$: Evaporation rate of the equilibrium
state for the DM mass range from $1.5$ to $3.0$ GeV, obtained from
simulation and calculation, respectively. $Right$: Evolution of
the evaporation rate for a 2 GeV DM from its initial captured state
to the equilibrium state, against a time scale of the solar age $t_{\odot}$.
See text for details. }

\end{figure}
Evaporation rates obtained from these two approaches are found quite
consistent in our study. For illustration in the left panel of fig.~\ref{figEvaporation-rate}
shown are the relevant evaporation rates for a benchmark cross section
$\sigma_{\chi e}=10^{-40}\,\mathrm{cm}^{2}$ and the DM mass ranging
from $1.5$ to $3.0$~GeV, with the blue solid line representing
the evaporation rate drawn from the simulation, and the red dashed
line corresponding to the calculated one. In the right panel of fig.~\ref{figEvaporation-rate},
we present the simulated evolution of the evaporation rate for a 2
GeV DM. The time scale is expressed in terms of the solar age $t_{\odot}$.
 From the right panel of fig.~\ref{figEvaporation-rate}, the
 thermalisation time $t_{\mathrm{th}}$ can also be determined once the evaporation rate is observed to reach its convergence. For illustration, we present the simulated thermalisation time for cross section $\sigma_{\chi e}=10^{-40}\,\mathrm{cm}^{2}$ in fig.~\ref{fig:thermTime}.

\begin{figure}
\begin{centering}
\includegraphics[scale=0.80]{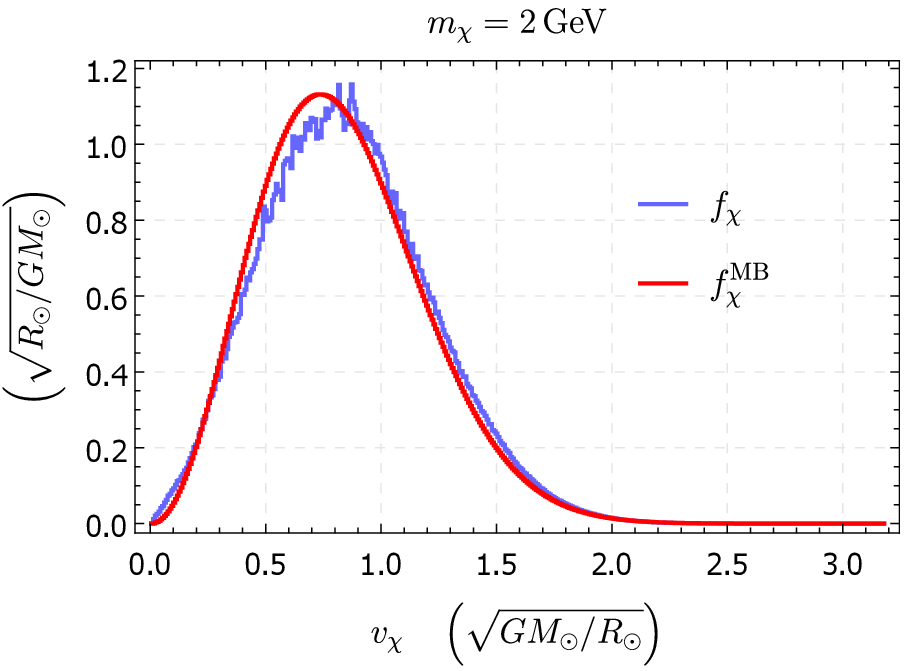}$\quad$\includegraphics[scale=0.57]{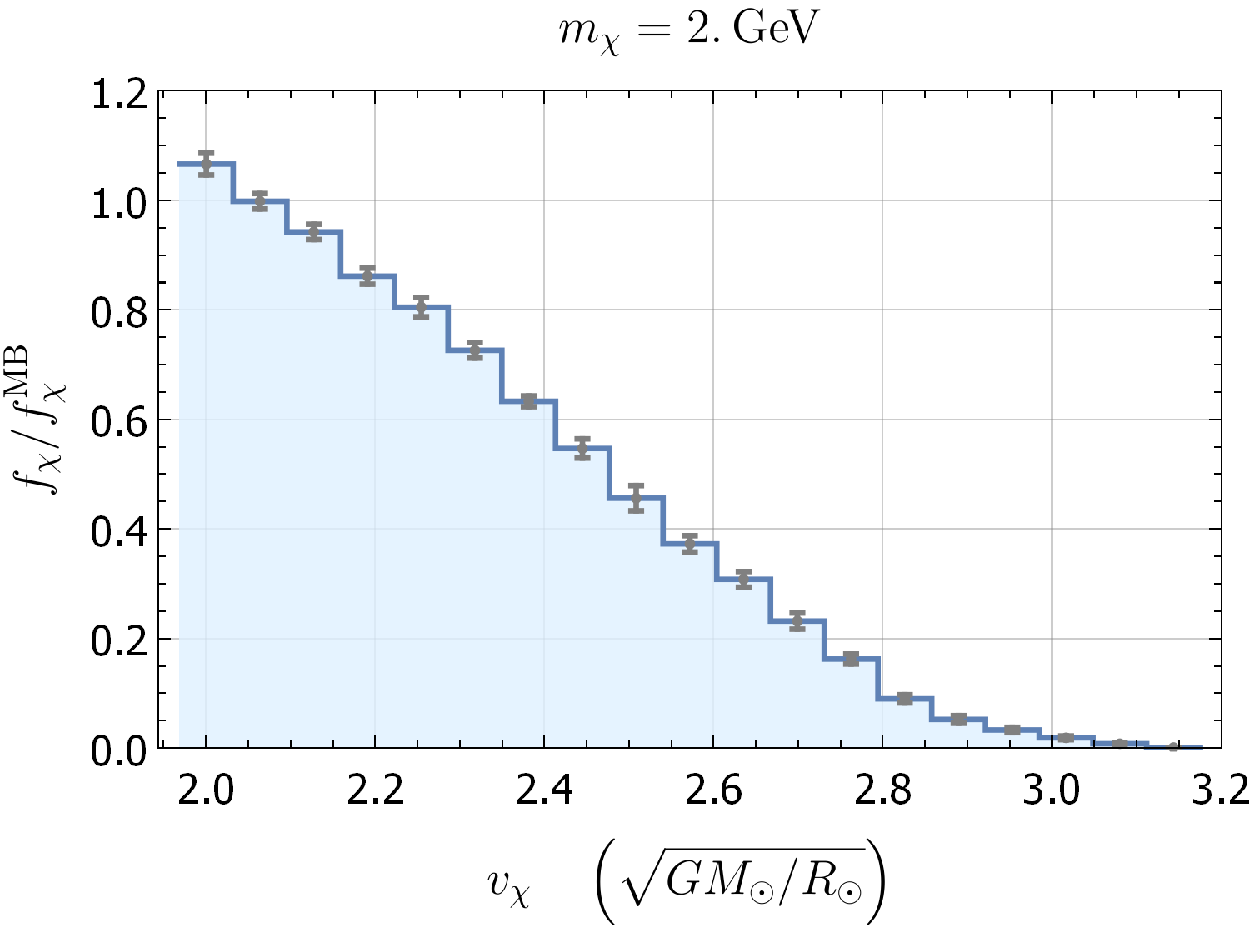}
\par\end{centering}

\protect\caption{\label{fig:VdistributionComparison}\textit{Left}: velocity distributions of
$2\,\mathrm{GeV}$ DM particle obtained from simulation (blue) and
Maxwell-Boltzmann approximation truncated at escape velocity with an effective temperature $T_{\chi}\approx0.845\, T_{\odot}\left(0\right)$
(red), respectively. The two distributions extend to no further than
the escape velocity at the solar core $v_{\mathrm{esc}}\left(0\right)\approx3.17$.
\textit{Right}: the ratio between the simulated and the isothermal velocity
distributions at the high velocity tail. See text for details.}

\end{figure}

It is interesting to compare above evaporation rate with the one given
in ref.~\cite{Garani:2017jcj}, where a Maxwell-Boltzmann form is
assumed for the solar DM distribution. By requiring that the net heat
transfer between the DM particles and the solar elements equal zero,
the effective temperature $T_{\chi}$ of the DM isothermal distribution
and hence the evaporation rate can be determined.
 In fig.~\ref{fig:VdistributionComparison} we present the
comparison between the two velocity distributions of a $2\,\mathrm{GeV}$
DM particle.   In the left panel, the simulated velocity distribution
$f_{\chi}$ is shown in the blue, while the approximated Maxwell-Boltzmann
form $f_{\chi}^{\mathrm{MB}}$ is shown in the red, with an effective
temperature $T_{\chi}\approx0.845\, T_{\odot}\left(0\right)$ drawn
from ref.~\cite{Garani:2017jcj}. While the bulks
of two velocity distributions truncated at the escape velocity are found to be basically consistent,
it turns out that the simulated ones fall off much more rapidly at
the high velocity end where evaporation substantially occurs, which
consequently results in greatly suppressed evaporation rates. As shown in the right panel, the
ratio between the two distributions, $f_{\chi}/f_{\chi}^{\mathrm{MB}}$, is significantly suppressed especially at the high velocity
tail, leading to an evaporation rate around 4 orders of magnitude
smaller. Such suppression has also been observed in the case of DM-nucleon
interaction in comparison between the two approaches~\cite{Nauenberg:1986em,Gould:1987ju,Liang:2016yjf},
but the extent is much slighter because in such case only a finite
part of evaporation events take place around the local escape velocity
where the suppression is remarkable. Considering that the evaporation rate is highly sensitive to tail of the velocity distribution, we perform three simulations to generate sufficient statistics to examine the robustness of our results. Owing to the fine grids  representative of the  high-energy bound states on the
$E$-$L$ plane, these simulated distributions at the high velocity end are found to be quite consistent, and  evaporation error is confined within  15\%,  corresponding to a variation in evaporation mass of 0.013~GeV \footnote[2]{\renewcommand{\baselinestretch}{1}\selectfont Conventionally, one used the evaporation mass $m_{\mathrm{evp}}$ defined through the equation $E_{\odot}\left(m_{\mathrm{evp}}\right)=t_{\odot}^{-1}
 $ as a rough estimate of the DM masses above which the evaporation effects can be neglected. It is straightforward to read from the slope in fig.~\ref{figEvaporation-rate} that a  variation of 15\% of evaporation rate translates to a displacement of the evaporation mass around 0.013~GeV.}. Thus we omit the statistical errors originating from the simulation in our discussion.

On the other hand, the annihilation coefficient $A_{\odot}$ is expressed
in terms of the thermal cross section $\left\langle \sigma v\right\rangle _{\odot}$
and the effective occupied volume of the solar DM, $V_{\mathrm{eff}}$,
as the following:
\begin{eqnarray}
A_{\odot} & \equiv & \frac{\left\langle \sigma v\right\rangle _{\odot}}{V_{\mathrm{eff}}},\label{eq:annihilationCoefficient}
\end{eqnarray}
if an $s$-wave thermal annihilation cross section is assumed. The
effective volume is defined as
\begin{eqnarray}
V_{\mathrm{eff}} & \equiv & \frac{\left(\int_{0}^{R_{\odot}}n_{\chi}\left(r\right)4\pi r^{2}\mathrm{d}r\right)^{2}}{\int_{0}^{R_{\odot}}n_{\chi}^{2}\left(r\right)4\pi r^{2}\mathrm{d}r},
\end{eqnarray}
with $n_{\chi}\left(r\right)$ being the number density of the solar
DM, which in practice is also determined from simulation. Therefore,
the effective volume can be approximated from the simulated equilibrium
distribution as the following function:
\begin{eqnarray}
V_{\mathrm{eff}} & = & 5.67\times10^{29}\,\left(\frac{5\,\mathrm{GeV}}{m_{\chi}}\right)^{2.15}\mathrm{cm}^{3}.
\end{eqnarray}

\section{\label{sec:MinimumTestableMass}Minimum testable mass of the leptophilic
DM }

Based on the above numerical efforts on the capture, evaporation and
annihilation of the leptophilic DM in the Sun, now we are ready to
explore the parameter space where the solar neutrino observational
approach is effective for the detection.

In analysis, we adopt the criterion $t_{\odot}/\tau_{e}\apprge3.0$
or equivalently $\tanh\left(t_{\odot}/\tau_{e}\right)\simeq1$ for
the assumption that the neutrino flux reaches its full strength. On
the other hand, in order to specify the parameter region for the annihilation-
and evaporation-dominated scenarios, we set the criteria as $E_{\odot}^{2}/\left(4A_{\odot}C_{\odot}\right)\leq0.1$
and $E_{\odot}^{2}/\left(4A_{\odot}C_{\odot}\right)\geq10$, respectively,
where the canonical $s$-wave thermal annihilation cross section $\left\langle \sigma v\right\rangle _{\odot}=3\times10^{-26}\,\mathrm{cm}^{3}\cdot\mathrm{s}^{-1}$
is adopted in definition~(see eq.~(\ref{eq:annihilationCoefficient})).

\begin{figure}
\begin{centering}
\includegraphics[scale=0.46]{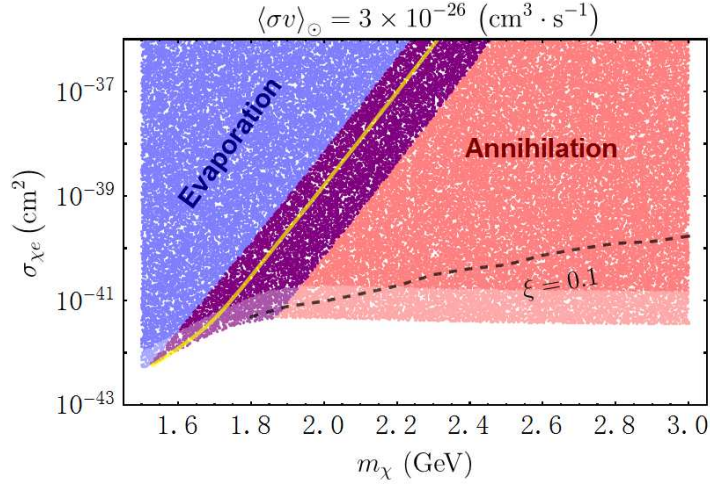}
\par\end{centering}

\protect\caption{\label{fig:parameterRegion}The parameter regions dependent on DM
mass $m_{\chi}$ and the DM-electron cross section $\sigma_{\chi e}$.
While the signal regions $\tanh\left(t_{\odot}/\tau_{e}\right)\simeq1$
are presented as the coloured areas, the lighter parts correspond
to the region where $0.9\leqslant\tanh\left(t_{\odot}/\tau_{e}\right)\apprle1$
for reference. In the red (blue) area, annihilation (evaporation)
plays the dominant role in the number evolution of the solar DM, and
the purple belt represents the transition zone between the two extreme
scenarios. The yellow solid line represents the evaporation mass defined in eq.~(\ref{othercriterion}). The black dashed line marks the contour  of the parameter
$\xi=0.1$. See text for details.}

\end{figure}

The relevant parameter regions are presented in fig.~\ref{fig:parameterRegion}.
The quantitative analysis enables us to draw clear boundaries among
different signal topologies. For instance, for a DM-electron cross
section $\sigma_{\chi e}=10^{-40}\,\mathrm{cm}^{2}$, the assumption
of the equilibrium between capture and annihilation is only valid
for a DM particle heavier than $1.94$~GeV, while for a DM mass smaller
than $1.72$~GeV, one can no longer extract the coupling strength
of the DM-electron interaction from the observed neutrino flux, because
the number of DM particles $N_{\chi}\simeq C_{\odot}/E_{\odot}$  turns independent
of cross section $\sigma_{\chi e}$. Moreover, if the cross section
$\sigma_{\chi e}$ is smaller roughly than $10^{-41}\,\mathrm{cm}^{2}$,
the equilibrium among capture, evaporation and annihilation has not
yet been reached at the present day. As a consequence, the signal
flux is suppressed and the unsaturated number of the solar DM particles
needs to be specified to determined or constrain the coupling strength~\cite{Albuquerque:2013xna}. For reference,  in  fig.~\ref{fig:parameterRegion} we also plot  in yellow solid line the evaporation mass from the definition adopted in refs.~\cite{Busoni:2013kaa,Garani:2017jcj}, where the minimum testable mass is defined as the one for which the number of captured DM particles differ with $C_{\odot}/E_{\odot}$ at the 10\% level
\begin{eqnarray}
\left|N_{\chi}-\frac{C_{\odot}}{E_{\odot}}\right|	=	0.1\, N_{\chi},\label{othercriterion}
\end{eqnarray} with $N_{\chi}$ defined in eq.~(\ref{eq:DMnumber}). It is evident that our definition of the evaporation mass is a little stricter than the above one.

In above investigation on the parameter space for the leptophilic
DM detection, the cross section is capped at $\sigma_{\chi e}=10^{-36}\,\mathrm{cm}^{2}$,
which corresponds to a mean free path $\ell_{\chi}\left(0\right)=\left(n_{e}\left(0\right)\sigma_{\chi e}\right)^{-1}\approx18\, R_{\odot}$
at the centre of the Sun. As coupling strength increases, collisions
between DM particles and electrons begin to be frequent in processes
such as capture, evaporation and energy transfer, and hence the optically
thin approximation will no longer be valid for the description of
the solar DM. Due to the multiple collisions, evaporation will be
suppressed and the minimum testable DM mass begin to decrease accordingly~\cite{Garani:2017jcj}.
In that regime, the Monte Carlo approach adopted in this study will
break down, since short DM free path requires an extra parameter for
the description of the solar DM distribution, as mentioned in Sec.~\ref{sub:sec.3.a}, and  a full consideration of the  Boltzmann equation is required, which is beyond the scope of this work.

Now we  make some final remarks on the methodology adopted in this work, namely, to what extent
is our calculation reliable considering that the DM depletion due
to annihilation is not included in simulation of the relaxation process.
To address this concern, we stress that it is the increment rather
than the deposit of solar DM particles that we are simulating, and
under this circumstance the annihilation effects can be all accounted
for by the DM particles that have already settled in the Sun. To verify
this, we first make a comparison between the  differential annihilation rate and the effective capture rate with respect to the relaxation time scale $t_{\mathrm{th}}$. During the time interval  $t_{\mathrm{th}}$, a number of $\Delta N_{\chi}\approx C_{\odot}\, t_{\mathrm{th}}$
DM particles are trapped  and participate in annihilation within the Sun,
with the effective capture rate $C_{\odot}$ involving the evaporation
effect. $\Delta N_{\chi}$ contributes approximately an annihilation
rate of $2A_{\odot}\left(t\right)N_{\chi}\left(t\right)\Delta N_{\chi}$,
where $N_{\chi}\left(t\right)$ is the number of the accumulated DM
particles and $A_{\odot}\left(t\right)$ is the annihilation coefficient
for corresponding equilibrium distribution at time $t$. It is noted
that in contrast to the relaxation time scale $t_{\mathrm{th}}$,
here $t$ should be regarded as a macroscopic temporal parameter.
This is no other than the instant thermalisation assumption, and its
validity will be scrutinised on a self-consistent basis in the following.
The actual annihilation rate is expected to be smaller and hence more
favourable to our reasoning because the initially captured DM particles
reside mostly in high orbits, where annihilation events are much more
rare than the case of equilibrium state. The accumulation of the solar
DM particles continues until the gains and losses reach a balance,
so one has $A_{\odot}\left(t\right)N_{\chi}\left(t\right)\Delta N_{\chi}\apprle A_{\odot}N_{\chi}^{2}\left(\Delta N_{\chi}/N_{\chi}\right)\apprle C_{\odot}\left(\Delta N_{\chi}/N_{\chi}\right)$,
with $N_{\chi}$ being the saturate number of the solar DM particles
and corresponding annihilation coefficient $A_{\odot}$. The only assumption introduced here is  that the $equilibrium$ distribution at $t$  approximately equals the one   corresponding to the saturated number $N_{\chi}$, so one has $A_{\odot}\left(t\right)\simeq A_{\odot}$. Its validity also relies on the instant thermalisation that  will be investigated later.  Above inequities indicates that, at the time scale of relaxation $t_{\mathrm{th}}$,  the annihilation rate due to the freshly captured DM particles can be well contained by the effective capture rate $C_{\odot}$. Especially,  if condition  $\Delta N_{\chi}/N_{\chi}\ll1$ is satisfied, the differential annihilation rate can be neglected compared to the effective capture rate.  In this case, the equation (now at a microscopic time scale) governing the evolution of the DM particles captured during time interval $t_{\mathrm{th}}$  regresses to a linear one, and the states of these DM particles at a microscopic time $\tau$ within $\left(0,\, t_{\mathrm{th}}\right)$ can be expressed as a linear superposition of states of  independent samples captured during an arbitrary time interval $\delta t\left(\ll t_{\mathrm{th}}\right)$, at different moments in the time sequence $\left\{ 0,\,\delta t,\,2\delta t,\cdots,\tau-\delta t,\, \min\left(\tau,\, t_{\mathrm{th}}\right)\right\}$.   To be specific, the equation to describe the DM number can
be written as
\begin{eqnarray}
\frac{\mathrm{d}N_{\chi}}{\mathrm{d}\tau} & = & \widetilde{C_{\odot}}-E_{\odot}\left(\tau\right)N_{\chi}\left(\tau\right),\label{eq:masterEq}
\end{eqnarray}
where the capture rate $\widetilde{C_{\odot}}$ does not include the
evaporation effects. By linearity,
one can re-express Eq.~(\ref{eq:masterEq})  as
\begin{eqnarray}
\frac{\mathrm{d}\left(N_{\chi0}+N_{\chi1}+\ldots N_{\chi n}\right)}{\mathrm{d}\tau}=\widetilde{C_{\odot}}-\left(E_{\odot0}\left(\tau\right)N_{\chi0}\left(\tau\right)+\ldots E_{\odot n}\left(\tau\right)N_{\chi n}\left(\tau\right)\right),
\end{eqnarray}
 or equivalently,
\begin{eqnarray}
\frac{\mathrm{d}N_{\chi,i}}{\mathrm{d}\tau} & = & -E_{\odot,i}\left(\tau\right)N_{\chi,i}\left(\tau\right)\nonumber \\
\nonumber \\
\frac{\mathrm{d}N_{\chi,0}}{\mathrm{d}\tau} & = & \widetilde{C_{\odot}}-E_{\odot0}\left(\tau\right)N_{\chi0}\left(\tau\right),\label{eq:0_partEq}
\end{eqnarray}
with $1\leq i\leq n=\lfloor\left(\tau/\delta t\right)\rfloor$, where the evaporation rate $E_{\odot,i}\left(\tau\right)$
only depends on the $i$-th DM particle sample.
In the optical thin regime, one can always take a small enough time interval $\delta t$ for which the post capture collisions are too soon to occur, so the supply of freshly captured DM particles dominates the evolution of  the  $0$-th sample, and Eq.~(\ref{eq:0_partEq}) can be simplified as
\begin{eqnarray}
\frac{\mathrm{d}N_{\chi,0}}{\mathrm{d}\tau} & = & \widetilde{C_{\odot}}.
\end{eqnarray}
Once the DM number of the $0$-th sample reaches the fixed value $N_{\chi,0}=\widetilde{C_{\odot}}\cdot\delta t$, we change its label to $n+1$, and use the following equation to describe
its state thereafter,
\begin{eqnarray}
\frac{\mathrm{d}N_{\chi,n+1}}{\mathrm{d}\tau} & = & -E_{\odot,n+1}\left(\tau\right)N_{\chi,n+1}\left(\tau\right),\label{eq:n+1 partEq}
\end{eqnarray}
where the evaporation rate $E_{\odot,n+1}\left(\tau\right)$ does
not depends on other samples of DM particles. Then we allocate one more
sample labeled $0$ to account for the capture effects and repeat above procedures.

Therefore, so far as the simulated DM increment is concerned, while the evaporation
effect of the volatile DM particles is taken into full account, the
annihilation effect can be reasonably neglected as long as the condition $\Delta N_{\chi}/N_{\chi}=C_{\odot}\, t_{\mathrm{th}}/N_{\chi}\simeq\xi\equiv C_{\odot}\, t_{\mathrm{th}}/\left[\sqrt{C_{\odot}/A_{\odot}+E_{\odot}^{2}/\left(4A_{\odot}^{2}\right)}-E_{\odot}/\left(2A_{\odot}\right)\right]\ll1$
is fulfilled, where the saturate number is approximated  as the analytic expression $N_{\chi}=C_{\odot}/\left(\sqrt{C_{\odot}A_{\odot}+E_{\odot}^{2}/4}+E_{\odot}/2\right)$ in eq.~(\ref{eq:DMnumber}). From an effective perspective the relaxation process can be conceived as a buffer, and only those settle  into the equilibrium state are considered as captured and subject to the ensuing annihilation.

Since the annihilation effect is not included
in the DM effective capture rate $C_{\odot}$, $N_{\chi}/C_{\odot}$
is smaller than actual saturation time of the increasing solar DM number. In addition, considering that the annihilation grows to its maximum at $N_{\chi}$ and the relation $\xi\geq A_{\odot}N_{\chi}^{2}\, t_{\mathrm{th}}/N_{\chi}$, $\xi$ also imposes a cap on the strength of the perturbation to the distribution due to annihilation. So the criterion $\xi\ll1$ also  applies  to the relaxation  in response to the annihilation for the accumulated DM particles and lives up to a sufficient condition for the instant
thermalisation. The instant thermalisation in turn validate the a priori assumption that $A_{\odot}\left(t\right)\simeq A_{\odot}$.  As a self-consistency check, the contour lines of the
ratio $\xi=0.1$ is given in fig.~\ref{fig:parameterRegion} for
the reference purpose. The parameter region above these contours corresponds
to smaller $\xi$. While in the evaporation-dominated regime the  the annihilation effects  can be safely ignored and evolution of solar DM particles can be well described by eq.~(\ref{eq:0_partEq}), the instant thermalisation proves
 to be a reasonable assumption for most of the parameter region in the annihilation-dominated area. This  justifies our treatment in which the DM particles are captured at an effective rate that includes only   evaporation effects during the instant relaxation process, and participate in  subsequent annihilation with the equilibrium distribution obtained from simulating the relaxation process.

\appendix

\begin{acknowledgments}
YLT is supported  by National Research Foundation of Korea (NRF) Research Grant NRF- 2015R1A2A1A05001869, and the Korea Research Fellowship Program through the National Research Foundation of Korea (NRF) funded by the Ministry of Science and ICT (2017H1D3A1A01014127).
\end{acknowledgments}

\begin{appendices}
\renewcommand\thesection{\Alph{section}}
\renewcommand{\theequation}{\Alph{section}.\arabic{equation}}

\section{\label{sec:appendixb}relative probability distribution of the transient
states}

In this appendix we will prove that for the given amount of the captured
DM particles, while suffer loss from evaporation, their relative probabilities
will eventually evolve to a steady distribution. Towards this end, we consider
a Markov process with finite discrete states, which is governed by
the equation
\begin{eqnarray}
\frac{\mathrm{d}p_{i}\left(t\right)}{\mathrm{d}t} & = & \sum_{j}\mathbb{W}_{ij}\, p_{j}\left(t\right),
\end{eqnarray}
or in a compact form
\begin{eqnarray}
\frac{\mathrm{d}p\left(t\right)}{\mathrm{d}t} & = & \mathbb{W}\, p\left(t\right),\label{eq:master equation2}
\end{eqnarray}
where $\mathbb{W}$ is the transition matrix that describes the gains
and losses between the (\textit{n}+1)-state probabilities $p_{i}\:\left(i=0,\,1,\,\cdots n\right)$.
Based on the following two properties of $\mathbb{W}$ :
\begin{flalign}
\mathbb{W}_{ij}\geq0, & \quad\mathrm{for}\: i\neq j;\nonumber \\
\sum_{i}\mathbb{W}_{ij}=0, & \quad\mathrm{for}\:\mathrm{each}\: j,
\end{flalign}
it is well known that regardless of the initial distribution $p\left(0\right)$,
the probabilities of the \textit{n}+1 states have a steady distribution
in the long-time limit~\cite{tagkey2007x}. On the other hand, eq.~(\ref{eq:master equation2})
has the following apparent solution,
\begin{eqnarray}
p\left(t\right) & = & e^{\mathbb{W}\, t}\, p\left(0\right).\label{eq:evolution solution}
\end{eqnarray}
While the transition matrix $\mathbb{W}$ may not be diagonalisable,
there exists an invertible matrix $\mathbb{S}$ almost doing the job
such that $\mathbb{S}^{-1}\mathbb{W}\,\mathbb{S}=\mathbb{J}$, where
the Jordan normal form is expressed as
\begin{eqnarray}
\mathbb{J} & = & \left(\begin{array}{cccc}
\mathbb{J}_{0} & 0 & \cdots & 0\\
0 & \mathbb{J}_{1} & \cdots & 0\\
\vdots & \vdots & \ddots & \vdots\\
0 & 0 & \cdots & \mathbb{J}_{m}
\end{array}\right),
\end{eqnarray}
with the $d_{\mu}$-dimensional Jordan block
\begin{flalign}
\mathbb{J}_{\mu}=\left(\begin{array}{ccccc}
\lambda_{\mu} & 1 & 0 & \cdots & 0\\
0 & \lambda_{\mu} & 1 & \cdots & 0\\
\vdots & \vdots & \ddots & \cdots & 0\\
0 & 0 & \cdots & \lambda_{\mu} & 1\\
0 & 0 & \cdots & 0 & \lambda_{\mu}
\end{array}\right),
\end{flalign}
and the sum over all dimensionalities of the blocks equals that of
the Jordan matrix, $i.e.$,
\begin{eqnarray}
\sum_{\mu}d_{\mu} & = & n+1.
\end{eqnarray}
Thus eq.~(\ref{eq:evolution solution}) can be further expressed
as
\begin{eqnarray}
p\left(t\right) & = & \mathbb{S}\, e^{\mathbb{S}^{-1}\mathbb{W}\,\mathbb{S}\, t}\,\mathbb{S}^{-1}\, p\left(0\right)\nonumber \\
 & = & \mathbb{S}\, e^{\mathbb{J}\, t}\,\mathbb{S}^{-1}\, p\left(0\right)\nonumber \\
\nonumber \\
 & = & \mathbb{S}\,\left(\begin{array}{cccc}
e^{\mathbb{J}_{0}t} & 0 & \cdots & 0\\
0 & e^{\mathbb{J}_{1}t} & \cdots & 0\\
\vdots & \vdots & \ddots & \vdots\\
0 & 0 & \cdots & e^{\mathbb{J}_{m}t}
\end{array}\right)\,\mathbb{S}^{-1}\, p\left(0\right)\nonumber \\
\nonumber \\
 & = & \mathbb{S}\,\left(\begin{array}{cccc}
e^{\lambda_{0}t}\cdot e^{\mathbb{Y}_{0}t} & 0 & \cdots & 0\\
0 & e^{\lambda_{1}t}\cdot e^{\mathbb{Y}_{1}t} & \cdots & 0\\
\vdots & \vdots & \ddots & \vdots\\
0 & 0 & \cdots & e^{\lambda_{m}t}\cdot e^{\mathbb{Y}_{m}t}
\end{array}\right)\,\mathbb{S}^{-1}\, p\left(0\right),
\end{eqnarray}
where
\begin{eqnarray}
e^{\mathbb{Y}_{\mu}t} & = & \exp\left[\mathbb{J}_{\mu}t-\mathrm{diag}\left(\lambda_{\mu},\,\lambda_{\mu}\cdots\lambda_{\mu}\right)t\right]\nonumber \\
\nonumber \\
 & = & \exp\left[\,\left(\begin{array}{ccccc}
0 & 1 & 0 & \cdots & 0\\
0 & 0 & 1 & \cdots & \vdots\\
\vdots & \vdots & \ddots & \cdots & 0\\
0 & 0 & \cdots & 0 & 1\\
0 & 0 & \cdots & 0 & 0
\end{array}\right)\cdot t\,\right]\nonumber \\
\nonumber \\
 & = & \left(\begin{array}{ccccc}
1 & t & \frac{t^{2}}{2!} & \cdots & \frac{t^{d_{\mu}-1}}{\left(d_{\mu}-1\right)!}\\
0 & 1 & t & \cdots & \frac{t^{d_{\mu}-2}}{\left(d_{\mu}-2\right)!}\\
\vdots & \vdots & \ddots & \cdots & \vdots\\
0 & 0 & \cdots & 1 & t\\
0 & 0 & \cdots & 0 & 1
\end{array}\right).
\end{eqnarray}
That is to say, the probability distribution $p_{i}\left(t\right)$
can be expressed as a linear combination of the terms $\left\{ e^{\lambda_{\mu}t}\cdot t^{n_{\mu}}\right\} $
$\left(0\leq\mu\leq m;\,0\leq n_{\mu}\leq d_{\mu}-1\right)$, which
proves our conclusion that the transient states still have a stationary
relative distribution in the long-time limit if their probabilities
are renormalised to a finite number, and the thermalisation time scale
can be described with the second largest real part of the set $\left\{ \lambda_{\mu}\right\} $.
\end{appendices}
\vspace{3cm}

\providecommand{\href}[2]{#2}\begingroup\raggedright\endgroup

\end{document}